\documentclass[12pt,journal,twoside,draftcls,onecolumn]{IEEEtran}

\usepackage{listings}

\usepackage{cite}
\usepackage{graphicx}
 \graphicspath{{figures/}}
 \DeclareGraphicsExtensions{.pdf,.jpeg,.png,.jpg,.eps}
 \usepackage{epstopdf}

\newcommand{\cnfct}{\textsc{chk}_{\mathrm{bin}}}
\newcommand{\bvec}[1]{\mathbf{#1}}
\newcommand{\E}{\mathbb{E}}
\newcommand{\atanh}{\operatorname{arctanh}}

\usepackage{array}
\usepackage[caption=false,font=footnotesize]{subfig}
\usepackage{fixltx2e}

\usepackage{url}

\usepackage{amsmath}
\usepackage{amssymb}

\usepackage{algorithm}
\usepackage[noend]{algpseudocode}

\ifCLASSOPTIONdraftcls
\usepackage{float}
\usepackage[nolists,nomarkers]{endfloat}
\newfloat{algorithm}{tbp}{loalgorithm}
\floatname{algorithm}{Algorithm}
\DeclareDelayedFloat{algorithm}{Algorithms}
\fi

\begin{document}
\title{Relaxed Half-Stochastic Belief Propagation}
%
%
%

\author{Fran\c{c}ois~Leduc-Primeau,~\IEEEmembership{Student~Member,~IEEE}, Saied~Hemati,~\IEEEmembership{Senior~Member,~IEEE}, Shie~Mannor,~\IEEEmembership{Senior~Member,~IEEE}, and Warren~J.~Gross,~\IEEEmembership{Senior~Member,~IEEE}%
\thanks{Manuscript submitted on \today. This work was supported by the Natural Sciences and Engineering Research Council of Canada (NSERC). A preliminary version of this work was presented at the IEEE Globecom 2009 conference.}
\thanks{Shie Mannor is with the Department of Electrical Engineering, Technion, Haifa 32000, Israel. Email: shie@ee.technion.ac.il. The other authors are with the Department of Electrical and Computer Engineering, McGill University, Montreal, QC H3A 2A7, Canada. Emails: francois.leduc-primeau@mail.mcgill.ca, \{saied.hemati, warren.gross\}@mcgill.ca.}
}

\markboth{Leduc-Primeau \MakeLowercase{\textit{et al.}}: Relaxed Half-Stochastic Belief Propagation}%
{IEEE Transactions on Communications}
%

\maketitle
\thispagestyle{empty}


\begin{abstract}
Low-density parity-check codes are attractive for high throughput applications because of their low decoding complexity per bit, but also because all the codeword bits can be decoded in parallel. However, achieving this in a circuit implementation is complicated by the number of wires required to exchange messages between processing nodes.
Decoding algorithms that exchange binary messages are interesting for fully-parallel implementations because they can reduce the number and the length of the wires, and increase logic density.
This paper introduces the Relaxed Half-Stochastic (RHS) decoding algorithm, a binary message belief propagation (BP) algorithm that achieves a coding gain comparable to the best known BP algorithms that use real-valued messages.
We derive the RHS algorithm by starting from the well-known Sum-Product algorithm, and then derive a low-complexity version suitable for circuit implementation.
We present extensive simulation results on two standardized codes having different rates and constructions, including low bit error rate results. These simulations show that RHS can be an advantageous replacement for the existing state-of-the-art decoding algorithms when targeting fully-parallel implementations.
\end{abstract}


\section{Introduction}
\label{sect_intro}

Low-density parity-check (LDPC) codes can approach channel capacity with a low decoding complexity per bit, making them attractive for a wide range of error correction applications. 
For most applications, the decoding operation must be performed by a custom circuit implementation because of the processing performance requirements. For a given coding gain, we seek to optimize the processing performance (throughput, latency) normalized to the circuit area.

The decoding of LDPC codes is part of a large family of problems that can be solved by iteratively passing messages in a \emph{factor graph} \cite{kschischang:2001}. When the desired decoding throughput is high, the most efficient implementation approach is to explicitly map the factor graph in hardware. This is known as a fully-parallel implementation. 
Because of the structure of LDPC factor graphs, the wiring complexity typically represents a large portion of the implementation complexity. It has a big impact on area requirements, as well as dynamic power consumption. Furthermore, the longer wires are likely to form critical paths and constrain the maximum clock frequency.
We will call a circuit graph a graph where a node corresponds to a localized circuit block, and an edge corresponds to a connection between two circuit blocks, composed of one or several wires. In the simplest fully-parallel implementation, the circuit graph is identical to the factor graph.

Different approaches have been proposed to reduce the wiring complexity of fully-parallel LDPC decoders. Circuit architectures have been proposed \cite{darabiha:2006,cushon:2010} that send messages serially on a single wire, at the expense of using a larger number of clock cycles to exchange messages. However, this does not change the topology of the circuit graph, which can still lead to routing congestion. In Split-Row decoders \cite{mohsenin:2010}, the topology of the circuit graph is modified by \emph{partitioning} check nodes into multiple sub-nodes that are then linked by two or four wires. The authors show this topology change can have a big impact on area and power requirements of the decoder. However, their approach suffers from increasing error rates as the number of partitions increases.
Finally, \emph{stochastic} decoding algorithms have been introduced as a low-complexity alternative to the standard algorithms. The technique was initially demonstrated only on small codes \cite{rapley:2003}, but since then stochastic decoding has been applied to longer LDPC codes used in communication standards \cite{sharifi-tehrani:2010}.
Since they rely on binary messages, stochastic algorithms have a low wiring complexity. They also use a very simple check node function, which can be partitioned arbitrarily without introducing any approximation. However, these stochastic algorithms suffer from a high latency, and an important loss in coding gain with respect to Sum-Product Algorithm (SPA) decoding.

In this paper, we introduce a new BP algorithm that is a binary message passing (BMP) algorithm, which for an LDPC decoder means that the algorithm is constrained to using only modulo-2 addition as the check node function. Hard-decision decoding algorithms such as ``Gallager-B'' and the stochastic algorithms mentioned previously are BMP algorithms.
The Relaxed Half-Stochastic (RHS) algorithm relies on stochastic messages for exchanging information between processing nodes, but differs from existing stochastic algorithms in the way messages are generated, and in the operations performed in the variable nodes. The name ``relaxed'' comes from the use in the variable node function of an estimation mechanism that is similar to the \emph{relaxation} step in a Successive Relaxation decoder \cite{hemati:2006a}.
Our results show that despite the constraint on the check node function, it is possible to achieve very good performance, in terms of error rate, latency, and throughput, while preserving a low implementation complexity.
The RHS algorithm can match or even outperform the error rate of BP algorithms that use real-valued messages, such as the Sum-Product algorithm.
Furthermore, an implementation only requires addition operations for the variable node function, XOR operations for the check node function, and two wires per (bi-directional) edge of the circuit graph. 

A preliminary version of the RHS algorithm was introduced in \cite{leduc-primeau:2009}. In this paper, we present several improvements that increase the performance while reducing complexity.
In Section~\ref{sect_background}, we briefly review the Sum-Product, Min-Sum and Normalized Min-Sum algorithms, and review the stochastic message representation.
In Section~\ref{sect_algorithm}, the algorithm is developed using the Sum-Product Algorithm as a basis. Following this, we show how the data throughput can be increased by changing the decoding rule at some pre-determined iterations, similar to \emph{Gear-Shift} decoding \cite{ardakani:2006}. We also present a method for lowering error floors that can be used with RHS or SPA decoders.
In Section~\ref{sect_implementation}, we derive a low-complexity implementation from the ideal case described previously.
Finally, in Section~\ref{sect_results}, we present simulation results for two standardized codes.
Results are presented for various versions of the algorithm, from the ideal case described in Section~\ref{sect_algorithm} to a version containing all the approximations required to achieve a low-complexity circuit implementation. The performance is described in terms of bit error rate, as well as average and maximum number of iterations.

\section{Background}
\label{sect_background}

An LDPC code can be modeled by a bipartite graph $\mathcal{G} = (\mathcal{V}, \mathcal{C}, \mathcal{E})$, where elements of $\mathcal{C}$, named \emph{check nodes} (CN), represent the parity-check equations, and elements of $\mathcal{V}$, named \emph{variable nodes} (VN), represent the variables of the parity-check equations. A variable can represent a transmitted symbol, or, in the case of punctured codes, an additional unknown that is not part of the transmitted information. An edge exists between a variable node and a check node if the variable is an argument of the parity-check equation associated with the check node. 
Throughout the paper, we denote the degree of a given variable node by $d_v$, and the degree of a given check node by $d_c$.

We will introduce Algorithm~\ref{alg_BP} as a template to be shared by all BP algorithms. This template will allow us to define all the algorithms in terms of a VN function $\textsc{var}$ and of a CN function $\textsc{chk}$, with subscripts indicating the algorithm.
Let $v_i \in \mathcal{V}$ denote the variable node with index $i \in [1,n]$, $c_j \in \mathcal{C}$ denote the check node with index $j \in [1,m]$, and $\mathcal{N}(x)$ denote the set of neighbors of a node $x$. In Alg.~\ref{alg_BP}, we use $V_i = \{j : c_j \in \mathcal{N}(v_i)\}$ and $C_j = \{i : v_i \in \mathcal{N}(c_j)\}$. We also denote a message from $v_i$ to $c_j$ by $\eta_{i,j}$ and from $c_j$ to $v_i$ by $\theta_{j,i}$. The $\textsc{init}(y_i)$ function converts a channel output $y_i$ into the representation domain used by the specific algorithm, and similarly $\delta$ is the value corresponding to a probability of $\frac{1}{2}$.
Finally, for a set $S = \{s_1, s_2, \ldots, s_z\}$, the expression $\textsc{var}(S)$ is equivalent to $\textsc{var}(s_1, s_2, \ldots, s_z)$, and similarly for $\textsc{chk}$.

In the paper we use the term \emph{throughput} to refer to the average number of bits processed by the decoder per time unit.
The BP algorithm terminates as soon as a codeword is found, and therefore in the discussions we will assume that the data throughput of the decoder is inversely proportional to the average number of iterations until convergence. We also use the term \emph{latency} to refer to the time required to run the decoder for $L$ iterations.

\begin{algorithm}
\caption{BP Template}\label{alg_BP}
\begin{algorithmic}
\Procedure{BP}{$\bvec{y}$}

	\State $\zeta_i \gets \textsc{init}(y_i)$, $\forall i$
	\State $\theta_{j,i} \gets \delta$, $\forall i, j$ 
	
	\For{$t \gets 1$ to $L$}
	
		\For{$i \gets 1$ to $n$}
			\ForAll{$j \in V_i$}
				\State $\eta_{i,j} \gets \textsc{var}( \{\zeta_i\} \cup \{ \theta_{a,i} : a \in V_i \} \setminus \{\theta_{j,i}\} )$
			\EndFor
		\EndFor
		\For{$j \gets 1$ to $m$}
			\ForAll{$i \in C_j$}
				\State $\theta_{j,i} \gets \textsc{chk}( \{ \eta_{a,j} : a \in C_j \} \setminus \{\eta_{i,j}\} )$
			\EndFor
		\EndFor		
		\State Compute the decision vector $\bvec{\hat{x}}(t)$
		\State Terminate if $\bvec{\hat{x}}(t)$ is a valid codeword
	\EndFor
	\State Declare a decoding failure
\EndProcedure
\end{algorithmic}
\end{algorithm}

\subsection{The Sum-Product Algorithm}
\label{sect_background_spa}

The Sum-Product algorithm is a BP algorithm that, for a cycle-free $\mathcal{G}$, computes the maximum-likelihood (ML) estimate of each codeword bit. 
Because of the cycles contained in an LDPC code's factor graph, the Sum-Product algorithm is not guaranteed to converge to the bit-wise ML estimates. Nonetheless, it has proven to be a very useful approximation algorithm.
The Sum-Product algorithm can be expressed in terms of various metrics, notably probability values or log likelihood ratios (LLR).
The \emph{log likelihood ratio} (LLR) metric is most often used for implementations because it reduces the quantization error,
and lowers the implementation complexity by avoiding the need for multiplications. An LLR metric $\Lambda$ is defined in terms of a probability $p$ as $\Lambda = \ln ((1-p)/p)$. Following \cite{kschischang:2001}, we will use the notation $\{\sim \!\! X_i\}$ to denote the set $\{X_1,X_2, \ldots , X_{n}\} \setminus \{X_i\}$, where $n$ is given implicitly from the context. 
In the LLR domain, a variable node output $\Lambda_{i}'$, $1 \leq i \leq d_v$, is given by
\begin{equation}
\label{eqn_spavnllr}
\ifCLASSOPTIONdraftcls 
\Lambda_{i}' = \textsc{var}_{\Lambda}(\sim \! \! \Lambda_i)
	= \Lambda_0 + \sum_{\Lambda_j \in \{\sim \Lambda_i \}} \Lambda_j
	= \Lambda_0 + \sum_{j=1}^{d_v} \Lambda_j  - \Lambda_i,
\else 
\begin{aligned}
\Lambda_{i}' =& \textsc{var}_{\Lambda}(\sim \! \! \Lambda_i) \\
	=& \Lambda_0 + \sum_{\Lambda_j \in \{\sim \Lambda_i \}} \Lambda_j \\
	=& \Lambda_0 + \sum_{j=1}^{d_v} \Lambda_j  - \Lambda_i,
\end{aligned}
\fi
\end{equation}
where $\Lambda_i$ is the input message on edge $i$ and $\Lambda_0$ is the a-priori likelihood obtained from the channel.
A common expression for computing a check node output $\Lambda_{i}'$, $1 \leq i \leq d_c$, is given by
\begin{equation}
\label{eqn_chkspa}
\ifCLASSOPTIONdraftcls 
\Lambda_{i}' = \textsc{chk}_{\Lambda}(\sim \! \! \Lambda_i)
	= \atanh \left( \prod_{\Lambda_j \in \{\sim \Lambda_i \}} \tanh(\Lambda_j) \right).
\else 
\begin{aligned}
\Lambda_{i}' =& \textsc{chk}_{\Lambda}(\sim \! \! \Lambda_i) \\
	=& \atanh \left( \prod_{\Lambda_j \in \{\sim \Lambda_i \}} \tanh(\Lambda_j) \right).
\end{aligned}
\fi
\end{equation}
However, the use of $\tanh$ and $\atanh$ is a source of quantization error in implementations.
The check node function in the Min-Sum algorithm \cite{fossorier:1999} removes the quantization error and reduces the complexity at the cost of an approximation. This check node function is given by
\begin{equation}
\label{eqn_chkms}
\Lambda_{i}' = \textsc{chk}_{\mathrm{MS}}(\sim \! \! \Lambda_i)
	= \min_{\Lambda_j \in \{\sim \Lambda_i \}} |\Lambda_j| \prod_{\Lambda_j \in \{\sim \Lambda_i \}} \mathrm{sign}(\Lambda_j),
\end{equation}
where
\ifCLASSOPTIONdraftcls
$\mathrm{sign}(x)=1$ if $x \geq 0$, and $\mathrm{sign}(x)=-1$ if $x < 0$.
\else 
\begin{equation*}
\mathrm{sign}(x)=
\begin{cases}
	1 & \text{if $x \geq 0$,} \\
	-1 & \text{otherwise.}
\end{cases}
\end{equation*}
\fi
The approximation can be improved by performing a multiplicative or additive correction on the $\min(\cdot)$ operation \cite{chen:2002,zhao:2005}. 
With a multiplicative correction, the algorithm is known as Normalized Min-Sum (NMS). The NMS check node function is given by
\begin{equation}
\label{eqn_nmschk}
\Lambda_{i}' = \alpha \, \textsc{chk}_{\mathrm{MS}}(\sim \! \! \Lambda_i),
\end{equation}
where $0 < \alpha \leq 1$ depends on the code structure. Although the optimal $\alpha$ should also depend on the channel signal-to-noise ratio (SNR), this can be ignored in practice \cite{chen:2002}.

\subsection{Stochastic Belief Representation}
\label{sect_background_stoch}

The stochastic stream representation expresses belief information by using a random sequence of binary messages, where the information is contained in the sequence's mean function.
In the probability domain, the Sum-Product check node function for $n$ inputs is given by \cite{gallager:1963}
\begin{equation}
\textsc{chk}_\mathrm{p}(p_1, p_2, ..., p_{n}) = \frac{1- \prod_{i=1}^{n} (1 - 2 p_{i})}{2}.
\end{equation}
For the stochastic check node function, we let the check node inputs be the random bit sequences $\{X_1(t), \ldots, X_n(t)\}$, independent and distributed such that $\E[X_i(t)]=p_i(t)$.
To evaluate the check node function on these stochastic stream inputs, we want the check node function binary output $Y(t)$ to be a random sequence with $\E[Y(t)] = \textsc{chk}_\mathrm{p}(p_1(t), ..., p_{n}(t))$.
This is satisfied by
\begin{equation}
\label{eqn_chkbin}
\ifCLASSOPTIONdraftcls
Y(t) = \cnfct(X_1(t), X_2(t), \ldots, X_n(t))
	= X_1(t) \oplus X_2 \oplus \ldots \oplus X_n(t),
\else 
\begin{aligned}
Y(t) =& \cnfct(X_1(t), X_2(t), \ldots, X_n(t)) \\
	=& X_1(t) \oplus X_2 \oplus \ldots \oplus X_n(t),
\end{aligned}
\fi
\end{equation}
where $\oplus$ represents modulo-2 addition.
Note that this is also the function used in Gallager's hard-decision decoding algorithms \cite{gallager:1963}.
The probability domain SPA variable node function for 2 inputs is
\begin{equation}
\label{eqn_probvn}
\textsc{var}_\mathrm{p}(p_1, p_2) = \frac{p_1 p_2}{(1-p_1)(1-p_2) + p_1 p_2}.
\end{equation}
The function can be obtained for more inputs by re-using the two-input function, e.g.~$\textsc{var}_\mathrm{p}(p_1,\allowbreak p_2,\allowbreak p_3) = \textsc{var}_\mathrm{p}(p_1, \textsc{var}_\mathrm{p}(p_2, p_3))$.
If we now let the stochastic variable node function inputs be $\{X_1(t), \ldots, X_n(t)\}$, we would like the binary output $Y(t)$ to be distributed such that $\E[Y(t)] = \textsc{var}_\mathrm{p}(p_1(t), \ldots, p_n(t))$. However, there is no memoryless binary-valued function that will achieve this.
Some functions with memory are proposed in \cite{rapley:2003,sharifi-tehrani:2010}.
In this paper, we introduce a new variable node function that is more accurate, and that can handle an extension of the concept of stochastic message to more than one bit.

\section{The RHS Algorithm}
\label{sect_algorithm}

The objective of the RHS algorithm is to achieve high-precision decoding while only relying on binary messages. The advantages of using binary messages include the smaller number of wires required for transmitting messages, and the low complexity and other interesting properties of the binary check node function (Eq.~(\ref{eqn_chkbin})), which will be discussed in Section~\ref{sect_chkproperties}.
Note that these advantages are not related to the number of bits that are sent on a given edge of the factor graph during one iteration, as long as the bits are sent sequentially on the same wire.
We can therefore introduce an extension of stochastic messages to sequences of binary messages. We will refer to these extended stochastic messages as ``iteration messages''.
In an RHS decoder, all information is exchanged between processing nodes in the form of stochastic messages, but the variable node computation can be performed in an other representation domain. This allows computing the variable node function with any desired accuracy.

\subsection{Check Node Function}
\label{sect_CNComputation}

As is the case for SPA, the RHS check node function takes as input $d_c - 1$ messages, and produces one output message. 
Ideally, for output $i$ this computation would be evaluating $m_i = \textsc{chk}_\mathrm{p}(\sim \!\! p_i)$.
However, the messages are constrained to be binary, and the CN instead evaluates (\ref{eqn_chkbin}) one or several times. Parameter $k>0$ controls the number of times (\ref{eqn_chkbin}) is used in one iteration.
We denote the binary inputs to the CN as $X_{i,j}$, where $i \in [1, d_c]$ is the input index, and $j \in [1, k]$ the bit index. 
The $j$-th binary check node evaluation for output $i$ is given by
\begin{equation}
\label{eqn_rhscn}
 Y_{i,j} = \cnfct(X_{1,j}, \ldots, X_{i-1,j}, X_{i+1,j}, \ldots, X_{d_c,j}).
\end{equation}
We will then define a function $g_k$ that estimates the ideal iteration message $m_i$ from the binary messages:
\begin{equation}
\label{eqn_msgestimator}
 \hat{m}_i = g_{k}(Y_{i,1}, Y_{i,2}, \ldots , Y_{i,k}).
\end{equation}
In practice, the variable node circuit receives the sequence $\{Y_{i,1}, Y_{i,2}, \ldots , Y_{i,k}\}$ and evaluates (\ref{eqn_msgestimator}), but conceptually, it belongs to the check node computation.
Note also that this message estimate corresponds to a single iteration of the algorithm, and should not be confused with the tracking estimator that will be introduced shortly.

We will denote by $\mathcal{M}$ the image of $g_k$, or the \emph{message set}. If the check node input messages $\{X_{1,j}, \ldots, X_{d_c,j}\}$ are independent and distributed such that $\E[X_{i,j}] = p_i$ (independent of $j$), the outputs $\{Y_{i,1}, Y_{i,2}, \ldots , Y_{i,k}\}$ are independent with $\E[Y_{i,j}] = m_i$, and the optimal message estimator $g_{k}$ is the sample mean:
\begin{equation}
\label{eqn_samplemean}
 \hat{m}_i = \frac{1}{k} \sum_{j=1}^{k} Y_{i,j}.
\end{equation}
Therefore the check node function is obtained by combining (\ref{eqn_samplemean}) and (\ref{eqn_rhscn}).
Under (\ref{eqn_samplemean}), $\mathcal{M}$ has $k+1$ elements. The $n$-th element of $\mathcal{M}$ will be denoted $\mu_n$, $0 \leq n \leq k$, and defined by $\mu_n = \hat{m}$ such that $\sum_{j=1}^k Y_{i,j} = n$.

\subsection{Variable Node Function}

\begin{figure}[tbp]
	\begin{center}
\ifCLASSOPTIONdraftcls
	\includegraphics[width=5in]{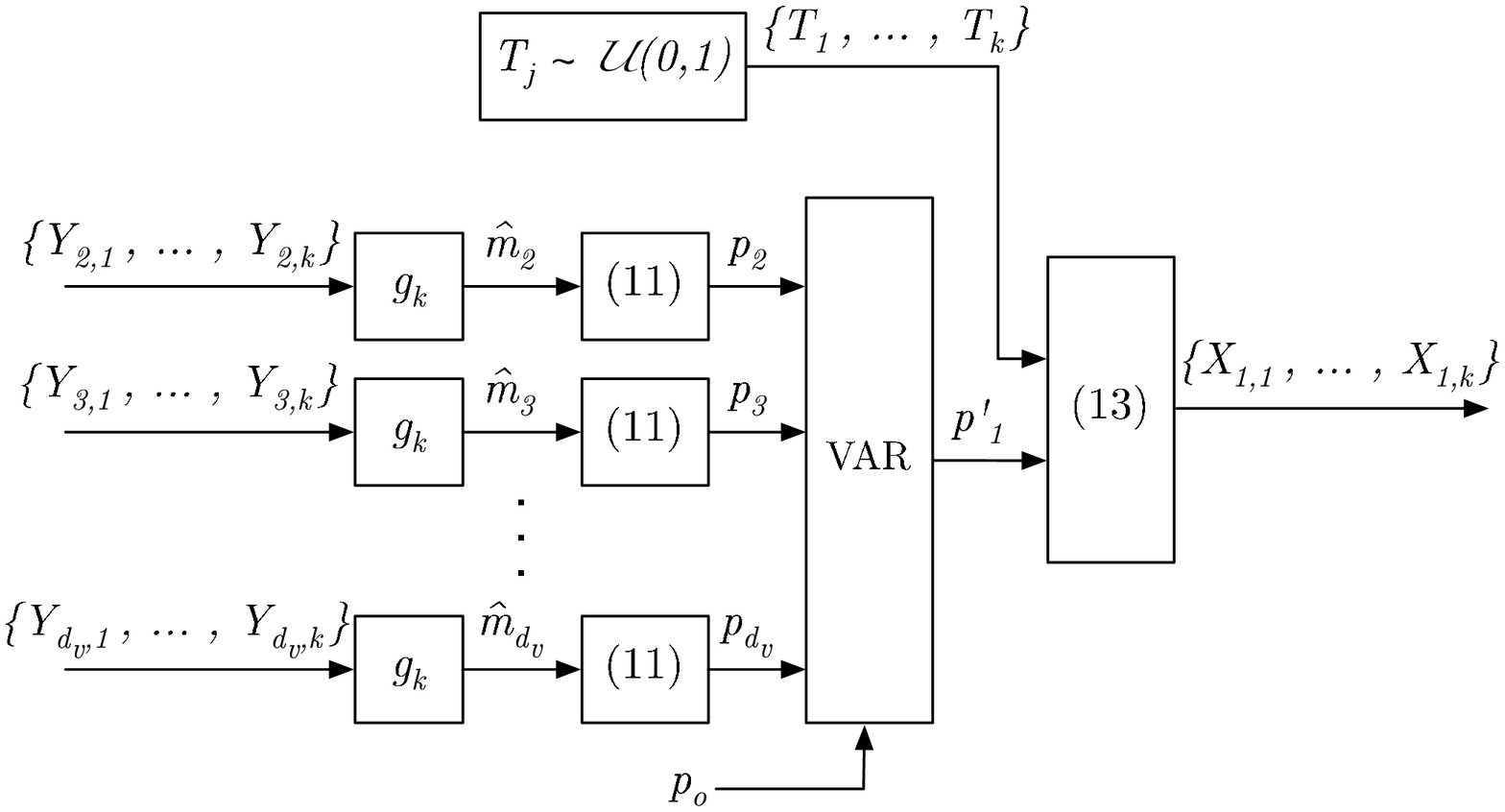}
\else
	\includegraphics[width=3.5in]{VN_functional_diag}
\fi
	\caption{Functional diagram of the RHS VN computation. For simplicity, only the output associated with edge $i=1$ is shown. An iteration index $t$ is implied for all variables, except the prior $p_o$. (\ref{eqn_vntracker}) and (\ref{eqn_binmsg}) refer to the respective equations.}
	\label{fig_vnfunct}
	\end{center}
	\end{figure}

	The variable node function for output $i$ takes as input $\{\sim \!\! \hat{m}_i\} \in \mathcal{M}^{d_v-1}$ and the codeword symbol prior $p_o$, and generates a binary output sequence $\{X_{i,1}, X_{i,2}, \ldots , X_{i,k}\}$. A functional representation of the computation is shown for a single output in Fig.~\ref{fig_vnfunct}.

	In the stochastic message-passing approach, the error correction performance of a decoder is decoupled from the precision of the messages exchanged by interpreting these messages as random sequences.
	In the RHS decoder, we are not interested in using the iteration messages $\hat{m}_i$ directly, but rather in extracting their underlying probability distributions $p_i(t)$.
	For the sake of simplicity, we use a linear tracking function, given by
	\begin{equation}
	\label{eqn_vntracker}
	p_i(t) = (1 - \beta) \, p_i(t-1) + \beta \, \hat{m}_i(t),
	\end{equation}
	where $\hat{m}_i(t)$ is the new received message, and $0 < \beta \leq 1$ is a real constant.
	
	Using the estimates $p_i(t)$ as input, the standard SPA function is then used to generate an intermediate output message $p_{i}'(t)$. The specific function used will depend on the representation domain. In the probability domain, the output probability for edge $i$, $p_{i}'(t)$, is given by
	\begin{equation}
	\label{eqn_SPAprob}
	p_{i}'(t) = \textsc{var}_\mathrm{p}(p_1(t), \ldots , p_{i-1}(t), p_{i+1}(t), \ldots , p_{d_v}(t), p_o).
	\end{equation}
	However, we will see in Section~\ref{sect_implementation} that the LLR domain is more convenient for the implementation.
	Finally, we generate the binary sequence that will be sent on edge $i$. This sequence is composed of $k$ independent binary random variables $X_{i,j}$ such that $\E[X_{i,j}] = p_{i}'(t)$, $1 \leq j \leq k$. This can be implemented by generating $k$ random \emph{thresholds} $T_j$, such that $T_j$ is uniformly distributed over $[0, 1]$, and constructing $X_{i,j}$ as
	\begin{equation}
	\label{eqn_binmsg}
	X_{i,j} =
	\begin{cases}
	1 & \text{if $p_{i}'(t) > T_j$,} \\
	0 & \text{otherwise.}
	\end{cases}
	\end{equation}

\subsection{Properties of the binary check node function}
\label{sect_chkproperties}

In an implementation of the RHS algorithm, the binary check node function (Eq.~(\ref{eqn_chkbin})) is the only operation performed on message bits as they are transmitted between variable node blocks.
Compared to the check node functions used in SPA or Min-Sum (Eq.~\ref{eqn_chkspa} to \ref{eqn_nmschk}), it has much lower complexity since only a modulo-2 sum is required. Furthermore, the various \emph{extrinsic} outputs of a check node can be computed in terms of a total function. For a given CN output $Y_i$ we have
\begin{equation}
\label{eqn_totalbinchk}
Y_i = \cnfct(\cnfct(X_1, X_2, \ldots , X_{d_c}), X_i).
\end{equation}
This can be used to simplify the implementation by computing only the total $Y_T = \cnfct(X_1,\allowbreak X_2,\allowbreak\ldots , X_{d_c})$ 
and broadcasting this result to all neighboring variable nodes, which then perform $Y_i = \cnfct(Y_T, X_i)$.
Finally, the most important property is that (\ref{eqn_totalbinchk}) can be factored arbitrarily. In a circuit implementation, this allows partitioning the logic in several locations, to provide flexibility in the circuit layout. In \cite{mohsenin:2010}, the ability to partition check nodes was shown to lead to large improvements in area, throughput and power. Contrary to \cite{mohsenin:2010}, check node partitioning in RHS requires no approximation and only uses two wires for linking partitions, since a message is transmitted on a single wire.
The circuit implementation aspects of check-node partitioning are discussed further in Section~\ref{sect_implementation}.

\subsection{$\beta$-sequences}
\label{sect_relaxation}

The choice of (\ref{eqn_vntracker}) as a tracking function makes RHS similar to a \emph{Successive Relaxation} SPA algorithm \cite{hemati:2006a}, but with the important difference that for RHS $\hat{m}_i$ and $p_i$ are defined on different domains.
Assuming that the maximum number of decoding iterations is fixed to $L$, we are interested in two performance metrics of the decoder, namely the bit error rate (BER) and the average number of iterations.
If we constrain $\beta$ to be constant, we can look for the value that optimizes some combination of these metrics. We will refer to this value as the optimal length-one $\beta$ sequence, denoted $\beta^{*}$. Experimental evidence shows that it depends on $k$, $L$, and on the channel SNR. This was also observed in \cite{hemati:2006a} for $L$ and SNR.
For a given $k$, $L$, and SNR, a simple way to identify $\beta^{*}$ is through Monte-Carlo simulation. Our results show that $\beta^{*}$ is only weakly affected by SNR, and that in practice this dependence can be ignored. Therefore the simulation can be performed at a moderate SNR, thereby ensuring that the computational complexity is reasonable. 
In our Monte-Carlo simulator, we included the ability to record BER as a function of the decoding iteration index. The BER can then be plotted in terms of the number of iterations, as in Fig.~\ref{fig_BERsettling} (we will call this ``settling curves''), or as a transfer function in the style of an error-probability EXIT chart \cite{ardakani:2004}, as in Fig.~\ref{fig_BERtransfer}.
By superimposing settling curves corresponding to different $\beta$ values, one can identify $\beta^{*}$ as a function of $L$, albeit with the constraint that $\beta^{*}$ is in the set of $\beta$ values simulated. 
For example, from Fig.~\ref{fig_BERsettling} we can determine that given $\beta^{*} \in \{0.5, 0.25, 0.15\}$, $\beta^{*} = 0.5$ for $L \leq 11$, and $\beta^{*} = 0.25$ for $11 < L \leq 50$. In this example, we have used a cost function that assigns a small but non-zero weight to the average number of iterations, such that the optimization is in terms of BER, but ties in BER are settled in favor of the faster algorithm.

	\begin{figure}[tbp]
	\begin{center}
\ifCLASSOPTIONdraftcls
	\subfloat[]{\label{fig_BERsettling}\includegraphics[width=4.5in]{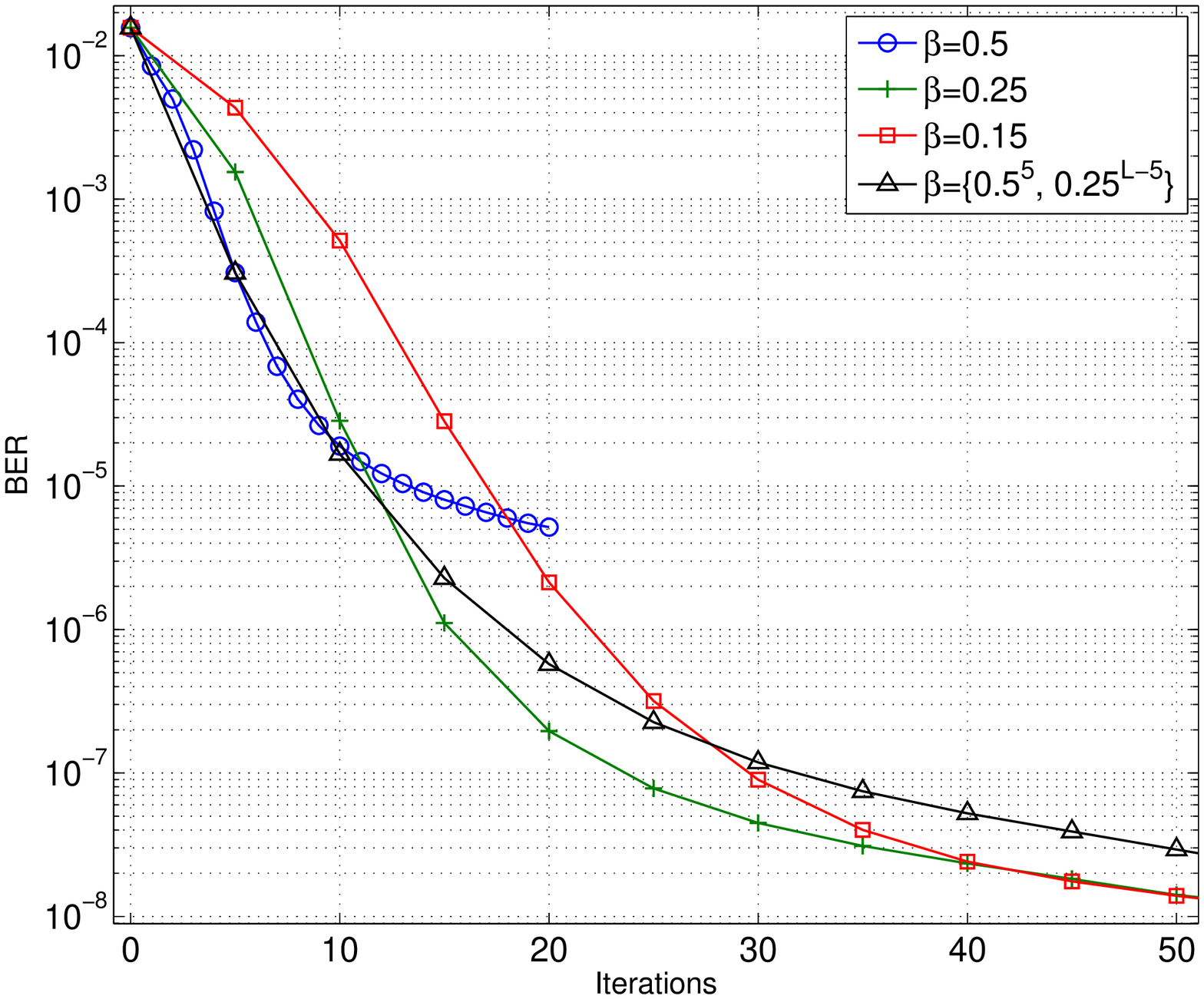}} \\
	\subfloat[]{\label{fig_BERtransfer}\includegraphics[width=4.5in]{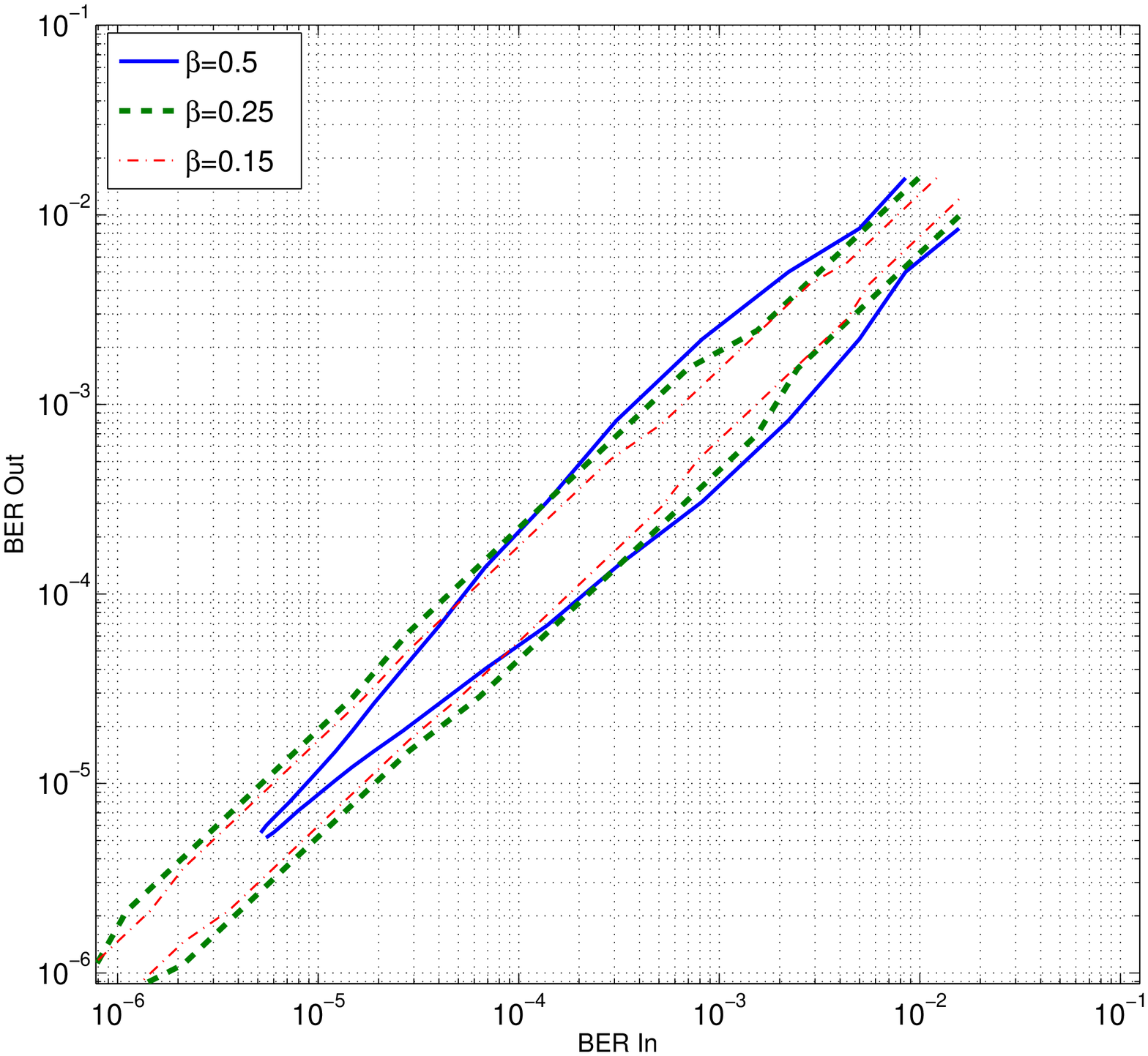}}
\else
	\subfloat[]{\label{fig_BERsettling}\includegraphics[width=3in]{10g_BERsettling_4p4dB}} \\
	\subfloat[]{\label{fig_BERtransfer}\includegraphics[width=3in]{10g_BERTransfer_4p4dB}}
\fi
	\caption{Bit error rate of RHS on the IEEE~802.3an code \cite{djurdjevic:2003} at 4.4~dB, (a) in terms of the number of decoding iterations, (b) in terms of the BER at the previous iteration, plotted in the style of an EXIT chart, with logarithmic axes.} 
	\end{center}
	\end{figure}

We now want to consider making $\beta$ a function of the iteration index $t$. A parallel can be made with Gear-Shift decoding \cite{ardakani:2006}, which considers how several decoding \emph{rules} can be used in sequence to optimize some performance metric (such as the maximum or the average number of iterations) while achieving a given target BER.
To find the optimal sequence of decoding rules, \cite{ardakani:2006} assumes that these rules have the following two properties. First, that the messages sent from variable nodes to check nodes can be described by a one-parameter probability density function (PDF). Let $c(t)$ be the PDF parameter at iteration $t$. The second necessary property is that $c(t) = f\left( c(t-1), c_o \right)$, where $c_o$ is the parameter of the channel output PDF.
Unfortunately, the second property does not hold for BP algorithms with memory. For example, in the case of RHS, the variable node input message $p_i(t)$ in (\ref{eqn_vntracker}) depends on all received messages $\{\hat{m}_i(1), \ldots, \hat{m}_i(t)\}$, which becomes clear when unrolling the recursion.

Our goal is to jointly optimize the BER and the average number of iterations.
As a design procedure, we propose to initially assume that the second property holds, i.e., that the algorithm is memoryless. 
In this case, the best $\beta(t)$ is simply the one that minimizes the BER at iteration $t$. Therefore the $\beta$-sequence can be read off the transfer function plot. Note that for a memoryless algorithm, the same sequence minimizes both the BER and the average number of iterations.
Since in reality the second property does not hold, this sequence is only used as a starting point. Following \cite{ardakani:2006}, we will write the sequences as $\{\beta_1^a, \beta_2^b, \ldots\}$ to mean that $\beta_1$ is used for the first $a$ iterations, followed by $\beta_2$ for the next $b$ iterations, and so on. If we constrain the sequence to be of the form $\{\beta_1^l, \beta_2^{L-l}\}$, only the parameter $l$ is left to be found, and if necessary it can be adjusted to trade off BER for throughput.
For example, Fig.~\ref{fig_BERtransfer} shows the BER transfer curves for various values of $\beta$. From the plot we see that $\beta=0.5$ is the best choice up to a BER-in of $3 \cdot 10^{-4}$, while for lower BER-in values, $\beta=0.25$ is superior. A BER of $3 \cdot 10^{-4}$ is achieved in 5 iterations with $\beta=0.5$, therefore the best sequence is $\{0.5^5, 0.25^{L-5}\}$. The actual BER performance of this sequence is shown in Fig.~\ref{fig_BERsettling}. Compared with the $\beta=0.25$ curve, the BER at 50 iterations is slightly degraded, but the average number of iterations is reduced by 33\%.
Note that for $L \leq 50$ (and most likely for any $L$), there are no length-one $\beta$-sequence that will yield this combination of BER and average number of iterations.

\subsection{Lowering the Error Floor}
\label{sect_floor}
A convenient way to improve the error floor at the level of the decoding algorithm is to consider a two-phase approach. In the first decoding phase, the normal algorithm is used. If there are unsatisfied check node constraints at the end of the first phase, the decoding is known to have failed and a modified algorithm is used to attempt to resolve the failure. If the two algorithms are similar, the cost in terms of circuit area is kept to a minimum. The two-phase approach has been widely used to improve error-floor performance (e.g.~\cite{varnica:2006,zhang:2008a,leduc-primeau:2010}).

The RHS algorithm should readily integrate with most Phase-II algorithms developed for SPA decoders since the RHS variable node operations are based on the SPA.
To illustrate this capability, we will introduce a Phase-II algorithm named ``VN Harmonization'' that can be used both for RHS and for SPA or Min-Sum decoders. Error rate results for this algorithm will be presented in Section~\ref{sect_results}.
VN Harmonization has some similarities with the Phase-II algorithm presented in \cite{zhang:2008a}, but has been found to be successful on certain codes for which the algorithm of \cite{zhang:2008a} is ineffective, such as the (2640,1320) Margulis code \cite{rosenthal:2000}. It also has the advantage that it operates locally in each variable node and requires no communication between processing nodes.

For each variable node and each iteration $t$, the VN Harmonization algorithm performs a modification on the set $I(t) = \{\Lambda_1(t), \ldots, \Lambda_{d_v}(t)\}$ of LLR-domain VN inputs. When used with the RHS algorithm, these LLR inputs correspond to the LLR-domain input trackers, which will be introduced in Section~\ref{sect_implementation::llrtrackers}.
We partition $I(t)$ into $I_{+}(t) = \{\Lambda \in I(t) : \Lambda \geq 0\}$, and $I_{-}(t) = I(t) \setminus I_{+}(t)$.
We then define a \emph{majority} set $M(t)$ as corresponding to the set with the largest number of elements among $I_{+}(t)$ and $I_{-}(t)$, with $M(t) = \emptyset$ if $|I_{+}(t)| = |I_{-}(t)|$, where $|S|$ denotes the cardinality of set $S$.
The algorithm is described by Algorithm~\ref{alg_VNH}, where $j$ is such that $M(t) = \{\Lambda_j(t)\}$ when $|M(t)|=1$, and $d \geq 0$ is a constant that must be found empirically.
\begin{algorithm}
\caption{VN Harmonization}\label{alg_VNH}
\begin{algorithmic}
\If {$|M(t)| = 1$}
	\For {$i \gets 1$ to $d_v$, $i \neq j$}
		\If {$\Lambda_j(t) \geq 0$}
			\State $\Lambda_i(t) \gets \Lambda_i(t) + d$
		\Else
			\State $\Lambda_i(t) \gets \Lambda_i(t) - d$
		\EndIf
	\EndFor
\EndIf
\end{algorithmic}
\end{algorithm}

\section{RHS Decoder Implementation}
\label{sect_implementation}

\begin{figure}[tbp]
\begin{center}
\ifCLASSOPTIONdraftcls
\includegraphics[width=5in]{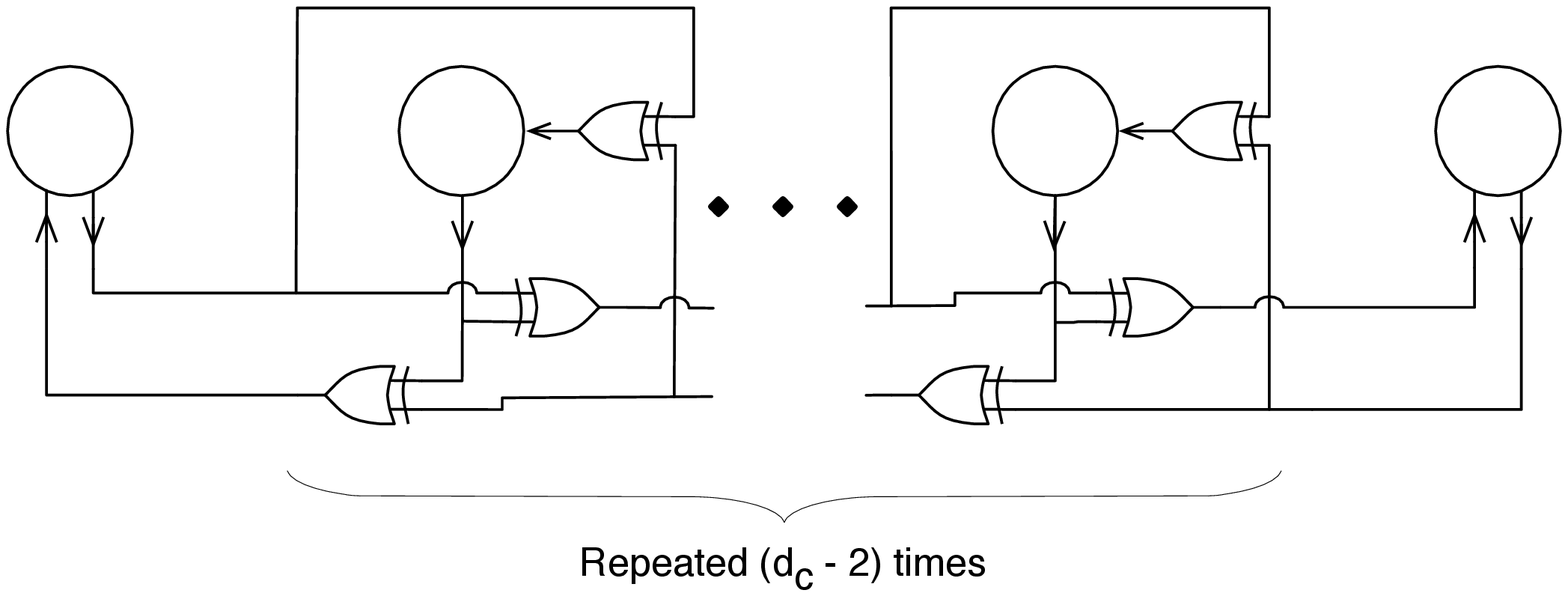}
\else
\includegraphics[width=3.4in]{fully_partitioned_chk}
\fi
\caption{Circuit diagram of a fully-partitioned check node. Circles represent variable node circuits.}
\label{fig_fullypartchk}
\end{center}
\end{figure}

We will now present an efficient implementation for the mechanisms introduced in the previous section.
By design, the RHS algorithm minimizes the wiring complexity of the decoder by using only one wire to transmit a given message. When $k>1$, the message bits are transmitted serially.
In addition, because of the properties of the check node function discussed in Section~\ref{sect_chkproperties}, the topology of the circuit graph can be modified to simplify wire routing. Interestingly, if a check node is fully partitioned, the associated logic can be integrated in the neighboring variable node blocks, eliminating the check nodes in the circuit graph. The circuit diagram for such a structure is shown in Figure~\ref{fig_fullypartchk}.
The degree of partitioning can be chosen individually for each check node, since it doesn't affect the RHS algorithm.
Since the check node function is very simple, the rest of this Section is devoted to the variable node function.
The LLR domain is attractive for implementing the VN function because (\ref{eqn_spavnllr}) is much simpler than the equivalent probability domain function. We now present an approach for implementing the RHS VN function in the LLR domain with low complexity.

	\subsection{Variable Node Output}
	
	For each edge $i$ of the variable node, $1 \leq i \leq d_v$, a sequence of random message bits $\{X_1, \ldots, X_k\}$ must be generated. This can be achieved by using (\ref{eqn_binmsg}), but we would now like to work with LLR values $\Lambda_{i}'$ instead of $p_{i}'$. As a result, the thresholds $T_j$ must be generated in the LLR domain.
	To generate LLR thresholds with low complexity, we approximate the natural logarithm with a base-2 logarithm, which can be generated using a simple circuit known as a \emph{priority encoder}. A priority encoder takes as input a binary sequence $\bvec{Z}$ and outputs the number $W$ of ``zero'' elements preceding the first ``one'' element in the sequence. If $\bvec{Z} = \{Z_1, \ldots, Z_q\}$ is generated by a sequence of independent Bernoulli experiments such that $\Pr(Z_i=1)=\psi_i$, the output $W \in \{0,1,\ldots,q\}$ has the following probability mass function:
	\begin{equation}
	\label{eqn_priorityPMF}
	\Pr(W=w) =
		\begin{cases}
		\psi_{w+1} \prod_{i=1}^{w} \left(1-\psi_i \right) & \text{$0 \leq w < q$,} \\
		\prod_{i=1}^{q} \left(1-\psi_i \right) & \text{$w=q$,} \\
		0 & \text{$w > q$.}
		\end{cases}
	\end{equation}
	The priority encoder only generates positive numbers, but since the LLR threshold distribution is symmetric, the sign bit can be generated separately using a fair random bit $S$.
	When quantized to integer values, the LLR threshold is expressed as $T_j = (-1)^S \, W$, for $W < q$. $\Pr(W=q)$ is a special case that is linearly related to $\Pr(W=q-1)$, and therefore the largest LLR magnitude that can be generated is $|T_j| = q-1$. When $W=q$, we can let $|T_j|$ take a value that is otherwise underrepresented by our approximation, e.g. let $T_j = (-1)^S \cdot 2$ if $W=q$.
	We now have to find $\{\psi_1, \psi_2, \ldots\}$ that best approximate the true LLR threshold magnitude distribution. If we consider again the case of an integer quantization, the probabilities $\psi_1=\frac{1}{4}$ and $\psi_2=\psi_3=\ldots=\psi_q=\frac{1}{2}$ provide a good approximation when $q$ is not too large. 
	In a circuit implementation, sequences of fair pseudo-random bits can be generated easily, for example using \emph{linear-feedback shift-register} circuits. $Z_1$ can be generated by combining two fair random bits, while $Z_2, \ldots, Z_q$ only require one fair random bit each.
		
	When deriving the stochastic check node function (\ref{eqn_chkbin}), we assumed that all binary messages entering a check node are statistically independent. Furthermore, in (\ref{eqn_samplemean}), we expect $\{Y_1, \ldots, Y_k\}$ to be independent, and therefore, in (\ref{eqn_binmsg}), the sequence $\{T_1, \ldots, T_k\}$ must be independent.
	To achieve this, the number $r$ of independent random numbers required in the entire decoder for one iteration is obviously at most $r = Nk$, where $N$ is the number of variable nodes.
	However, by relaxing the requirement for independence, the decoder can function with much less random numbers. For the codes considered in this paper, simulations have shown that a random number generator can be shared among 64 VNs without any degradation in performance, that is $r = Nk/64$. 
	As a result, the circuit implementation can contain only a few Random Number Generation modules, and the circuit area occupied by these modules is expected to be negligible.
	
	In practice, the LLR values $\Lambda_{i}'$ are represented on a finite range, and we must consider the impact this has on the operation of the decoder. Let this range be $[-\Lambda_\mathrm{cap}, \Lambda_\mathrm{cap}]$, with $\Lambda_\mathrm{cap} > 0$. If $|\Lambda_{i}'| \leq \Lambda_\mathrm{cap}$, the finite range has no impact. Therefore let $s$ be the number of check node inputs for which $|\Lambda_{i}'| > \Lambda_\mathrm{cap}$, $0 \leq s < d_c$. We can show that this approximately has the effect of changing the mean of $Y_j$, defined in (\ref{eqn_rhscn}), such that
	\begin{equation}
	\label{eqn_chkprobscaling}
	\E[Y_j] = \phi (m - \frac{1}{2}) + \frac{1}{2},
	\end{equation}
	where $\phi$ is a scaling factor that depends on $\Lambda_\mathrm{cap}$ and on $s$:
	\begin{equation}
	\label{eqn_phi}
	\phi = \left( 1 - \frac{2}{e^{\Lambda_\mathrm{cap}}+1} \right)^{s}.
	\end{equation}
	A consequence of (\ref{eqn_chkprobscaling}) is that, when $s>0$, the message estimator $g_k$ is no longer unbiased as defined in (\ref{eqn_samplemean}), and this has an impact on the error rate performance. To have $\E[\hat{m}]=m$, we must replace it with
	\begin{equation}
	\label{eqn_msgestimator2}
	\hat{m} = g_k(Y_1, \ldots, Y_k) = \frac{\sum_{j=1}^{k} Y_j}{k\phi} - \frac{1}{2\phi} + \frac{1}{2}.
	\end{equation}
	This in turn changes the message set $\mathcal{M}$, which is no longer a subset of $[0, 1]$. Similarly the codomain of the tracking function (\ref{eqn_vntracker}) would no longer be $[0,1]$ and it must be re-defined with the appropriate saturation. If we assume that $\phi \geq 1 - \frac{2}{k}$, we have $0 \leq \mu_n \leq 1$ for $n=\{1, \ldots, k-1\}$, and we can define the new tracking function as
	\begin{equation}
	\label{eqn_probtracker2}
	p_i(t) =
		\begin{cases}
		0 \text{, if $\hat{m}_i(t) = \mu_0$ and $p_i(t-1) < L$} \\
		1 \text{, if $\hat{m}_i(t) = \mu_{k}$ and $p_i(t-1) > H$} \\
		(1 - \beta) p_i(t-1) + \beta \hat{m}_i(t) \text{, otherwise,}
		\end{cases}
	\end{equation}
	where $L = \frac{\beta}{1 - \beta} \left( \frac{1}{2\phi} - \frac{1}{2} \right)$ and $H = 1 - L$.
	
	Ideally, the parameter $s$ in (\ref{eqn_phi}) would be set to the expected number of saturated check node inputs, which depends on the SNR and on the iteration index. However, having to set $s$ dynamically would make the decoder too complex, and we will resort to simple heuristic rules.
	At high SNR, a large portion of messages in a BP decoder become saturated\footnote{See \cite{varnica:2007} and \cite{schlegel:2010} for some observations on message saturation in LDPC decoding.}. Therefore, it seems reasonable to use $s = d_c-1$. Furthermore, when the $d_c$ values are small and $\Lambda_\mathrm{cap}$ is not too large, simulation results presented in Section~\ref{sect_results} show that such an $n$ can be used at all SNRs with almost no performance degradation. However, when the $d_c$ values are large, simulation results show significant variations in performance between $s=0$ and $s=d_c-1$, and a choice must be made based on the application.

	\subsection{Variable Node Input} 
	\label{sect_implementation::llrtrackers}
	
	\begin{figure}[tbp]
	\begin{center}
\ifCLASSOPTIONdraftcls
	\includegraphics[width=6in]{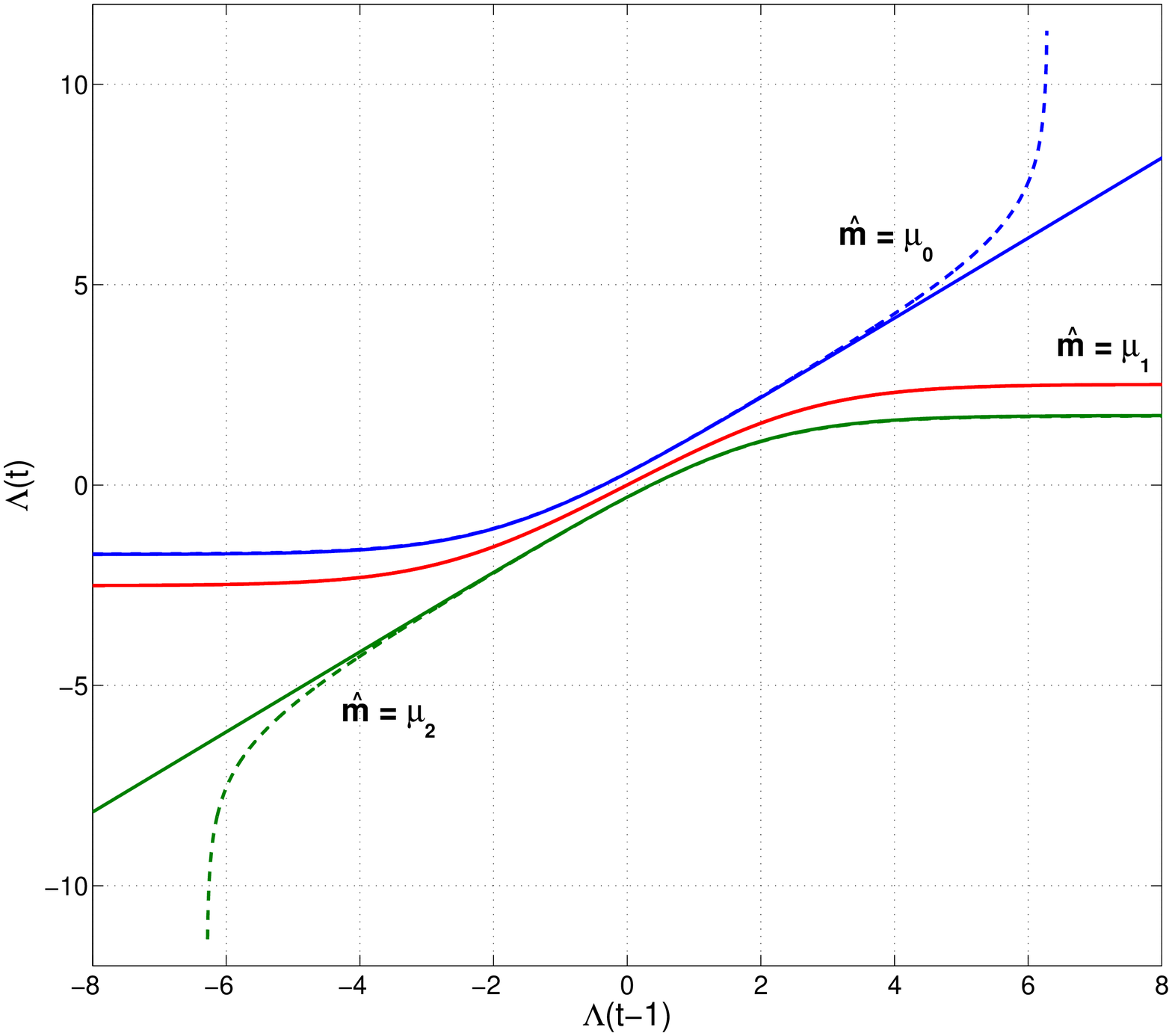}
\else
	\includegraphics[width=3.4in]{llrtracker_transfer_combined}
\fi
	\caption{Plot of (\ref{eqn_llrtracker}) with $\beta=0.15$ and $k=2$. Solid curves show the case of $\hat{m}(t) \in \{0, \frac{1}{2}, 1\}$ (i.e.~$s=0$). Dashed curves show the case of $s=31$ and $\Lambda_\mathrm{cap}=8$, with (\ref{eqn_msgestimator2}) as the message estimator.}
	\label{fig_llrtracker}
	\end{center}
	\end{figure}

	At the input of the variable node circuit, the values required to evaluate (\ref{eqn_spavnllr}) are estimated from the message stream using (\ref{eqn_vntracker}) (or (\ref{eqn_probtracker2})). However, since we choose to perform the variable node computations in the LLR domain, the estimate would need to be converted from the probability domain to LLR.
	There is a more interesting alternative, which is to design a tracking mechanism that operates directly in the LLR domain. 
	We will first consider how to achieve this when the tracking function is (\ref{eqn_vntracker}), and will later comment on how the mechanism should be modified if (\ref{eqn_probtracker2}) is used instead. Equation~(\ref{eqn_vntracker}) in the LLR domain becomes
	\begin{equation}
	\label{eqn_llrtracker}
	\Lambda(t) = \ln \left(
		\frac{e^{\Lambda(t-1)} + \beta (1 - \hat{m}(t) (e^{\Lambda(t-1)}+1)) }
		        {1 - \beta (1 - \hat{m}(t) (e^{\Lambda(t-1)}+1)) } \right).
	\end{equation}
	By fixing the message $\hat{m} \in \mathcal{M}$, we obtain a transfer function that describes the tracker update, which we denote $f(\Lambda; \hat{m})$. The tracker update is then expressed as
	$\Lambda(t) = f(\Lambda(t-1); \hat{m})$.
	With the message estimator given by (\ref{eqn_samplemean}) or (\ref{eqn_msgestimator2}), the transfer functions have the following symmetry property: $f(\Lambda; \mu_j) = -f(-\Lambda; \mu_{k-j})$, for $\frac{k}{2} < j \leq k$. An example is shown in Fig.~\ref{fig_llrtracker}.
	Therefore, the number of transfer functions we need to consider is $\lfloor k / 2 \rfloor + 1$. The remaining functions are simply obtained (and implemented) using the symmetry property.
	
	The transfer functions are non-linear in the LLR domain, but we need the tracking circuits to have a low complexity. Fortunately, the transfer functions can suitably be approximated by a linear function, combined with saturation functions at either or both ends of the linear domain. For each value of $\hat{m}$, the steps for deriving the simplified transfer functions are as follows:
	\begin{enumerate}
	\item Determine the image $A$ of $f(\Lambda; \hat{m})$. When $A$ is open-ended, that is for some $c \in \mathbb{R}$, $A=[-c, \infty$ or $A=-\infty, c]$, we restrict it to a finite interval $A=[-c, \Lambda_L]$ or $A=[-\Lambda_L, c]$. $\Lambda_L > 0$ is the maximum absolute value that must be represented in the trackers. It depends on the code structure and is the same as the maximum value that must be represented in an SPA decoder.
	\item For $\Lambda \in A$, find the optimal linear approximation $a \Lambda + b$ to $f(\Lambda; \hat{m})$.
	\item To simplify the circuit implementation, we want $a$ and $b$ to have binary representations that are as compact as possible, and therefore the constants are rounded according to this criterion. When possible we prefer $a=1$. This step requires some simulation of the decoder to determine how much the constants can be rounded.
	\end{enumerate}
	
	\emph{Example 1:}
	Fig.~\ref{fig_llrtracker} shows the transfer functions of the LLR-domain estimator for the case of 2-bit messages (i.e. $k=2$ in (\ref{eqn_msgestimator}) and (\ref{eqn_samplemean})) and $\beta=0.15$. In this case, the possible message estimates are $\mathcal{M} = \{0, \frac{1}{2}, 1\}$.
	For $f(\Lambda; 0)$, $A=[-1.73, \Lambda_L]$. We notice that the slope of $f(\Lambda; 0)$ on this range is close to 1, therefore we look for an approximation of the form $f(\Lambda, 0) = \Lambda + b$. 
	With $\Lambda_L=15$, the mean squared error is minimized by $b=0.206$, which we round to $b=\frac{1}{4}$. $A$ is rounded to $[-\frac{7}{4}, 15]$.
	$f(\Lambda; 1)$ is obtained by symmetry, and for $\hat{m}=\frac{1}{2}$, we get $f(\Lambda; \frac{1}{2}) = 0.776 \, \Lambda$ on the domain $[-2.5,2.5]$, which we round to $f(\Lambda; \frac{1}{2}) = \frac{3}{4} \Lambda$.
	
	Ultimately, the tracking circuit can be very simple. In the example above, the tracker value can be represented on 7 bits, and the only operations that are required are addition with $\pm1$ (representing $\pm \frac{1}{4}$), and multiplication by $\frac{3}{4}$, which can be implemented as the addition of two shifted values.
	
	When using $s>0$, the transfer functions are similar, except that $f(\Lambda; \mu_0)$ goes to infinity at the LLR value corresponding to $p(t-1)=L$, and similarly $f(\Lambda; \mu_k)$ goes to negative infinity at the LLR value corresponding to $p(t-1)=H$, as shown in Fig.~\ref{fig_llrtracker}. The maximum value to be represented in the tracker will therefore be set to $\Lambda_L = \ln \frac{1-L}{L}$ ($\Lambda_L=6.29$ in our example). However, for $f(\Lambda; \mu_0)$ and $f(\Lambda; \mu_k)$, using range $A$ described above as the linear approximation domain results in a poor approximation, since the functions are highly non-linear near $\Lambda_L$ and $-\Lambda_L$, respectively. The domain of the linear approximation must be reduced slightly in order to obtain a good fit.
	If we consider the quantized representation of $\Lambda(t)$, the ``infinity'' values can be handled by simply assigning a special meaning to the largest positive and negative values. We then re-define (\ref{eqn_spavnllr}) to take into account this special meaning. We first define
	a saturation indicator $S_i$ as $S_i = 1$ if $\Lambda_i = \Lambda_L$, $S_i = -1$ if $\Lambda_i = -\Lambda_L$, and $S_i=0$ if $|\Lambda_i| < \Lambda_L$.
	When $\sum_{S_j \in \{\sim S_i\}} S_j = 0$, the output $\Lambda_{i}'$ is given by (\ref{eqn_spavnllr}) as usual. Otherwise, if $\sum_{S_j \in \{\sim S_i\}} S_j > 0$, we set $\Lambda_{i}' = \Lambda_\mathrm{cap}$, and if $\sum_{S_j \in \{\sim S_i\}} S_j < 0$, $\Lambda_{i}' = -\Lambda_\mathrm{cap}$.
	
	The proposed linear approximation of the LLR-domain estimator can also support an efficient implementation of the $\beta$-sequences introduced in Sect.~\ref{sect_relaxation}. In the case of the example above, using multiple values for $b$ was found to provide a throughput advantage comparable to using $\beta$-sequences in a decoder that uses (\ref{eqn_vntracker}) directly.

\section{Simulation Results}
\label{sect_results}

To test the performance of the RHS algorithm we consider two standardized codes. The first is taken from the IEEE~802.3an standard. The code structure is based on a shortened Reed-Solomon code (RS-LDPC) \cite{djurdjevic:2003}. The code has length $2048$, rate $0.8413$, and is regular with $d_v=6$ and $d_c=32$. The second code has been standardized 
in \cite{ccsds:2007}. 
It has length $2048$, with an additional $512$ punctured variable nodes, rate $1/2$, and $d_v \in \{1,2,3,6\}$, $d_c \in \{3,6\}$.
The code design is known as Accumulate-Repeat-4-Jagged-Accumulate (AR4JA) \cite{divsalar:2005}.

\begin{figure}[tbp]
\begin{center}
\ifCLASSOPTIONdraftcls
	\includegraphics[width=6in]{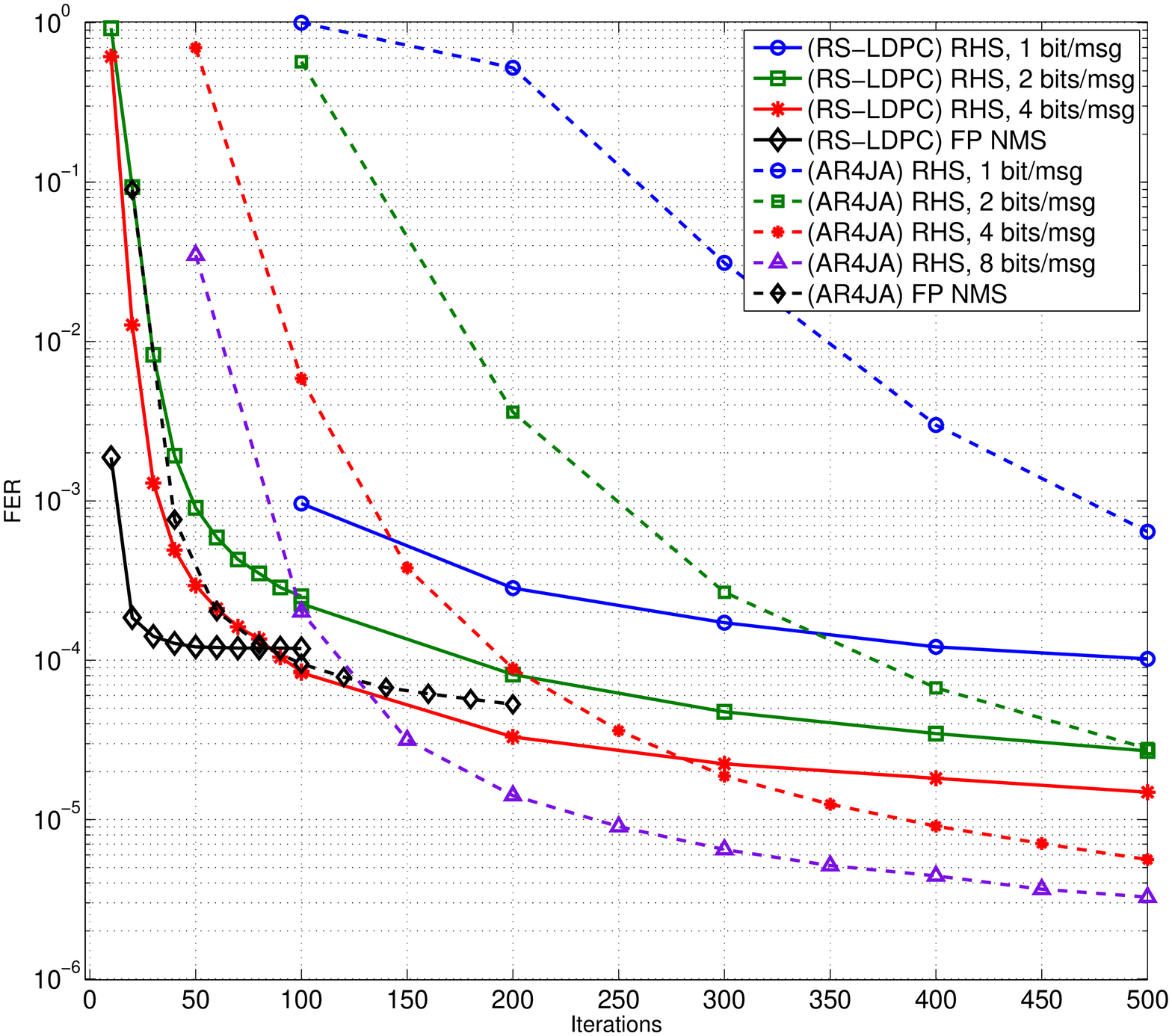}
\else
	\includegraphics[width=3.7in]{settling_rsldpc4dB_ar4ja2dB}
\fi
\caption{Frame error rate in terms of the maximum number of iterations. Solid curves are for the RS-LDPC code, at 4~dB, and dashed curves are for the AR4JA code, at 2~dB. The RHS algorithm is simulated with floating-point probability tracking (Eq.~\ref{eqn_vntracker}) and various message precisions, and compared with floating-point NMS.}
\label{fig_settlingAll}
\end{center}
\end{figure}

The performance of the various decoding algorithms is measured using software Monte-Carlo simulations executed on a parallel computing platform.
Following the discussion in Sections~\ref{sect_algorithm} and \ref{sect_implementation}, we will present results for various levels of idealization of the variable node tracking functions. We will refer to the use of (\ref{eqn_vntracker}) as ``floating-point (FP) probability tracking''. The second step towards implementation is to use linearized LLR-domain tracking (as described in \ref{sect_implementation::llrtrackers}) with full-precision parameters, referred to as ``optimal linear tracking''. Note that even in this case, we constrain the $\mu_0$ and $\mu_k$ solutions to be of the form $\Lambda(t) = \Lambda(t-1) + b$, such that the corresponding circuit is simply an adder. The last step is to round the linear tracking parameters to reduce the circuit implementation complexity. This is referred to as ``rounded linear tracking''. For the IEEE~802.3an code, we went further and implemented the trackers with integer data types and quantized channel outputs, to mimic the operation of a circuit implementation.

An RHS decoder has two design parameters. First, the iteration limit $L$ should be chosen based on the latency requirement. The other parameter is the number $k$ of bits per message. Figure~\ref{fig_settlingAll} shows the effect of $k$ on the frame error rate for the two codes simulated. A larger $k$ improves the error rate for a given $L$. Since the number of pre-determined linear operations that must be supported in the tracker circuit is $\lfloor k/2 \rfloor + 1$, the implementation complexity will generally grow with $k$. 
However, $\beta^{*}$ and the corresponding $b$ are increasing functions of $k$, and a larger $b$ can have the effect of decreasing the number of quantization bits required in the LLR tracker.
Since message bits are transmitted serially, a larger $k$ also increases the circuit latency associated with message transmission, but for small values of $k$, the variable node circuit latency is expected to be dominant.
Parameter $\beta$ is optimized as described in \ref{sect_relaxation}, with BER as the optimization target (unless mentioned otherwise). The variable node output LLR range, $\Lambda_\mathrm{cap}$, is chosen as small as possible. We use $\Lambda_\mathrm{cap}=8$ for the IEEE~802.3an code, and $\Lambda_\mathrm{cap}=6$ for the AR4JA code. 
We will first discuss in Section \ref{sect_results::maxBER} the maximum BER performance of the RHS algorithm, that is the BER when $L$ is chosen such that any increase provides a negligible BER improvement. Then, in Section \ref{sect_results::practical}, we consider how RHS performs when the focus is on decoding latency and throughput.

\subsection{Error Correction Capability}
\label{sect_results::maxBER}

\begin{figure}[tbp]
\begin{center}
\ifCLASSOPTIONdraftcls
	\includegraphics[width=6in]{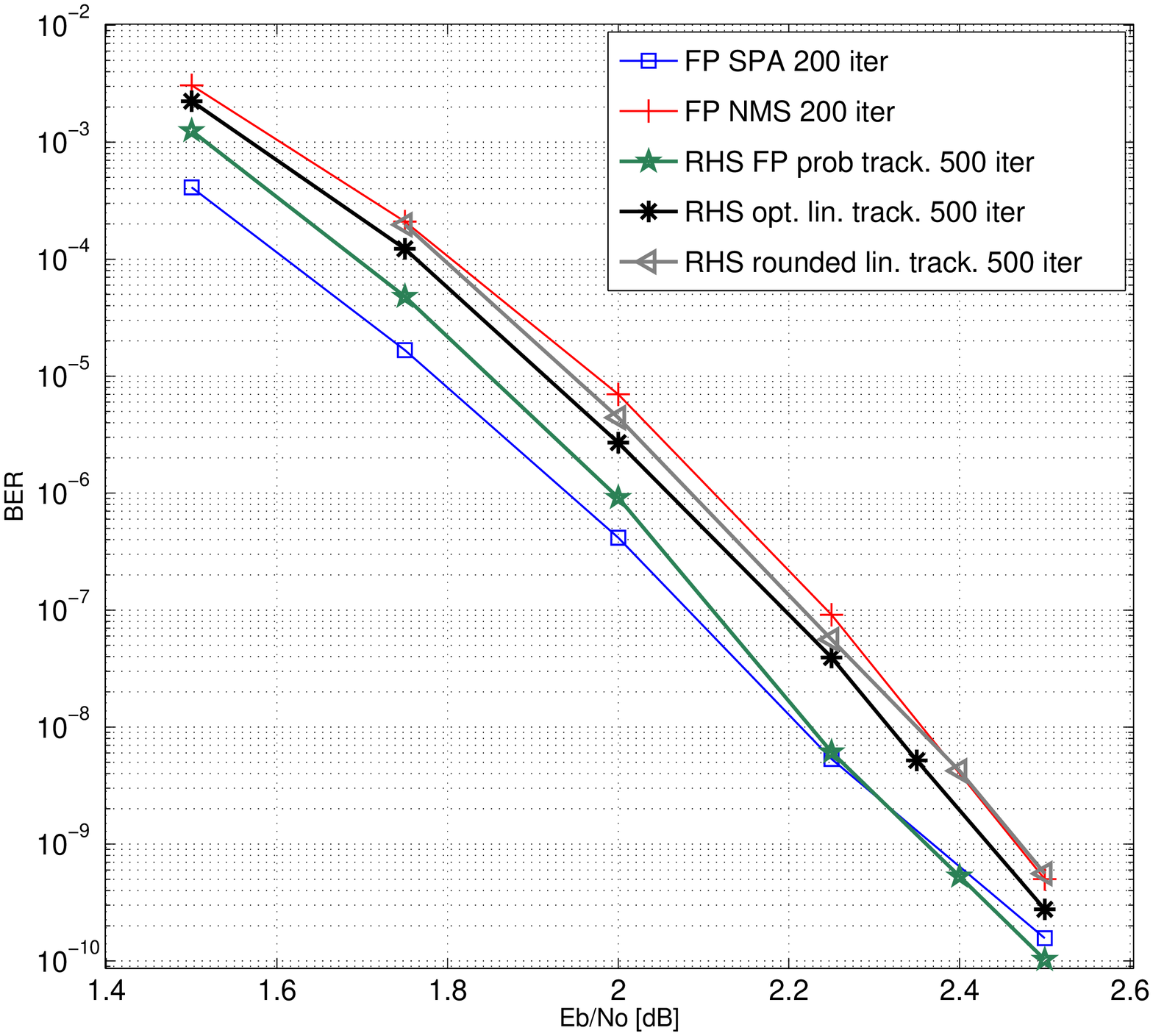}
\else
	\includegraphics[width=3.6in]{ar4ja1_BER_8}
\fi
\caption{Bit error rate performance on the AR4JA code. RHS uses $k=4$ bits per message.}
\label{fig_BERar4ja}
\end{center}
\end{figure}

The BER results are shown in Fig.~\ref{fig_BER10g_A} and \ref{fig_BER10g_B} for the RS-LDPC code, and in Fig.~\ref{fig_BERar4ja} for the AR4JA code. We show the error rate achieved by floating-point SPA and NMS as a reference. For NMS, parameter $\alpha$ in (\ref{eqn_nmschk}) is set to $\alpha=0.5$ for the RS-LDPC code, and $\alpha=0.75$ for the AR4JA code. 
As seen in the figures, RHS can match the performance of FP SPA, but also outperforms it by a significant margin on the RS-LDPC code. 
This superior performance on the RS-LDPC code can be attributed to the successive relaxation iterative dynamics that are a consequence of (\ref{eqn_vntracker}). In fact, we have verified that the RHS curve with 1K iterations and FP probability tracking shown in Fig.~\ref{fig_BER10g_A} exactly matches the BER of Successive Relaxation SPA \cite{hemati:2006a}.

Both codes considered are affected by message saturation effects at low error rates, which can cause error floors. For the RS-LDPC code, the connection between the error floor and message saturation is well documented in \cite{schlegel:2010}. For the AR4JA code, we have observed that when decoding with FP NMS, enforcing a limit on LLR messages causes a floor. For example, when limiting LLR values to the range $[-16, 16]$, the BER of FP NMS never goes below $10^{-9}$. Without the saturation limit, no floor is observed.
Because of the message saturation effects, for the AR4JA code we use (\ref{eqn_probtracker2}) as the basis for variable node message tracking, with $s=d_c-1$ in (\ref{eqn_phi}). In the waterfall SNR region, we have observed little difference in BER between $s=0$ and $s=d_c-1$, which motivates the use of the latter. We can see in Fig.~\ref{fig_BERar4ja} that no floor is observed on either FP NMS or RHS.
For the RS-LDPC code, RHS has an error floor that is comparable with quantized NMS. On this code, we have a solution available to address floor performance that has low complexity and is very effective, namely the VN Harmonization algorithm that was introduced in \ref{sect_floor}. This solution is therefore preferred over the use of $s>0$, especially since we have observed that for the RS-LDPC code, using $s>0$ degrades the BER in the waterfall region. We present some curves that use a decoding Phase-II with VN Harmonization at specific SNRs, denoted by (*), (**) and (***) in Fig.~\ref{fig_BER10g_A} and \ref{fig_BER10g_B}. The parameter $d$ in Alg.~\ref{alg_VNH} is set to $0.3$. For the other RHS curves in Fig.~\ref{fig_BER10g_A}, no Phase-II is used because the resulting BER would require too much computing time to be simulated.

As expected, the best BER performance is obtained when using (\ref{eqn_vntracker}) implemented with floating-point operations. We first want to observe the impact of the linear approximation to LLR-domain tracking. We can see that for both codes, the ``optimal linear tracking'' curves are close to FP probability tracking. 
We then consider the BER performance when low-complexity parameters are used for the LLR-domain tracking. To give an example of how we can expect the algorithm to perform once implemented in hardware, we show a simulation of the RS-LDPC code where the software implementation uses integer data types and the channel outputs are quantized on 4 bits. The specific parameters used were presented in the example of Section~\ref{sect_implementation::llrtrackers}. The BER achieved by this simulation (``rounded linear tracking'') shows approximately a 0.1~dB gain over NMS with 4-bit inputs, similar to the difference observed between ``optimal linear tracking'' RHS and FP NMS.
On the AR4JA code, the ``rounded linear tracking'' curve uses the same software implementation as ``optimal linear tracking'', but with rounded parameters. For the AR4JA code, the BER curves are shown for $k=4$, and therefore 3 different linear calculations must be implemented in the tracker. 
Since we choose $s=d_c-1$ in (\ref{eqn_phi}), $f(\Lambda; \hat{m})$ will depend on the degree $d_c$ of the check node generating the message $\hat{m}$. 
The AR4JA code has $d_c \in \{3, 6\}$. However, after rounding the parameters, the transfer functions are the same for both check node degrees, with the exception that $\Lambda_L = 6.25$ for $d_c=3$, and $\Lambda_L = 5.5$ for $d_c=6$. The transfer functions used for the ``rounded linear tracking'' result are as follows:
$f(\Lambda; \mu_0) = \Lambda + \frac{1}{2}$, $-1 \leq \Lambda \leq \Lambda_L$;
$f(\Lambda; \mu_1) = \frac{3}{4} \Lambda + \frac{1}{4}$, $-2 \leq \Lambda \leq \frac{11}{4}$;
$f(\Lambda; \mu_2) = \frac{1}{2} \Lambda$, $-2 \leq \Lambda \leq 2$.
The functions are saturated outside the ranges specified. As was the case for the RS-LDPC code, these tracking functions have a low implementation complexity. Furthermore, $\Lambda(t)$ can be represented on only 6~bits.

\begin{figure}[tbp]
\begin{center}
\ifCLASSOPTIONdraftcls
	\subfloat[]{\label{fig_BER10g_A}\includegraphics[width=4.4in]{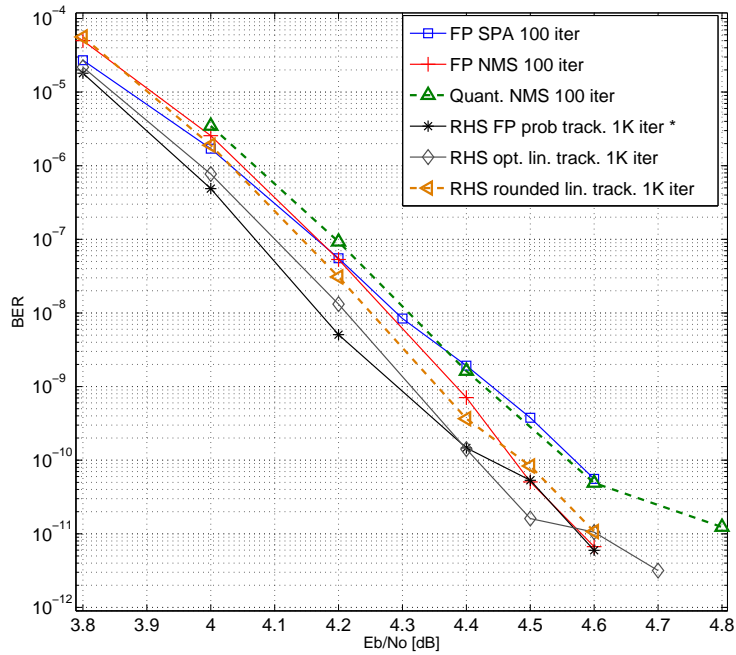}} \\
	\subfloat[]{\label{fig_BER10g_B}\includegraphics[width=4.4in]{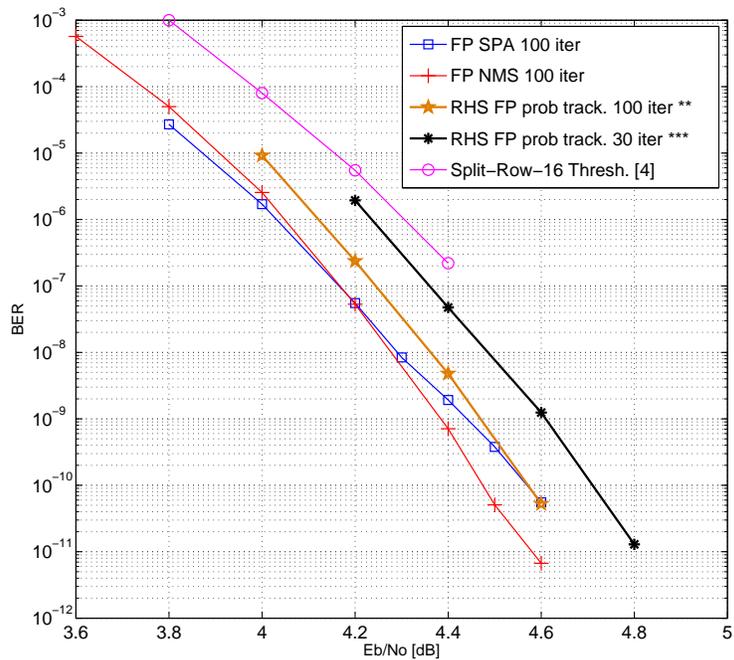}}
\else
	\subfloat[]{\label{fig_BER10g_A}\includegraphics[width=3.6in]{rs-ldpc_BER_A}} \\
	\subfloat[]{\label{fig_BER10g_B}\includegraphics[width=3.6in]{rs-ldpc_BER_B_2}}
\fi
\caption{{Bit error rate performance on the RS-LDPC code. Results for quantized decoders are shown with dashed curves. All RHS curves use $k=2$ bits per message. (*) At 4.6dB a 50-iteration Phase-II is used after Phase-I. (**) At 4.6dB a 25-iteration Phase-II is used. (***) At 4.8~dB a 30-iteration Phase-II is used.
The \emph{Split-Row} curve \cite{mohsenin:2010} uses 16 partitions and 4-wire links between partitions to handle the Threshold mechanism, and 11 iterations.}}
\end{center}
\end{figure}

\subsection{Throughput and Latency}
\label{sect_results::practical}

If we consider the iteration limit required for RHS to achieve the same performance as Normalized Min-Sum, we expect a higher number of iterations to be needed for RHS, because more information is exchanged in one iteration of NMS. 
The relationship between a number of iterations and actual time depends on the circuit latency associated with one iteration. This in turn depends on the amount of sequential logic involved in computing the messages, and on the wire delays, which often cannot be neglected in fully-parallel implementations. Furthermore, the metric of interest is usually not simply the latency, but the latency normalized to circuit area.
Because of its very simple check node update operation, RHS has less sequential logic per iteration. Furthermore, its ability to be partitioned arbitrarily improves the wire delays and area requirements of the circuit implementation.
With NMS, check nodes cannot be easily partitioned to improve the circuit implementation efficiency. An approach for doing so is proposed in \cite{mohsenin:2010}, but it has the effect of degrading the decoding performance. The curve taken from \cite{mohsenin:2010} appearing in Fig.~\ref{fig_BER10g_B} shows the BER when the check nodes are split into 16 partitions, with a limit of 11 iterations and 5 bits per message. The figure shows that better performance is achieved by RHS with a limit of 30 iterations, that is, with a comparable number of bits being exchanged.
Additionally, the BER performance is not tied to check node partitioning. For example, the implementation can feature fully-partitioned check nodes (layout as in Fig.~\ref{fig_fullypartchk}) while achieving a BER that is about three orders of magnitude better than the ``Split-Row-16'' curve, if a latency of 1000 iterations can be tolerated.

As presented in \ref{sect_relaxation}, the average number of iterations required for convergence can be reduced by using multiple $\beta$ values in sequence. 
For the RS-LDPC code, we found that despite exchanging only 2-bit messages, and despite the simple check nodes, the average number of iterations can be reduced to 3.46 (at 4.6~dB) by using $\beta = \{0.5^5, 0.25^{L-5}\}$. The RHS curve with $L=100$ iterations in Fig.~\ref{fig_BER10g_B} uses this $\beta$ sequence. In comparison, the NMS algorithm requires an average of 2.5 iterations at 4.6~dB, and therefore RHS requires only 38\% more iterations.
Using the same code, the authors of \cite{mohsenin:2010} have reported a 3.3$\times$ improvement in throughput from the use of check node partitioning. In addition, RHS can exchange less bits per iteration, and the check node circuits are much simpler. Therefore, in a fully-parallel circuit implementation, we expect the RHS algorithm to provide a higher nominal throughput (without considering area requirements) than NMS on this code. Considering throughput normalized to circuit area should further increase the advantage of RHS.
On the AR4JA code, we have observed that using $\beta=\{0.5^{26}, 0.25^{L-26}\}$ provides a 28\% throughput improvement (at 2.5~dB) over $\beta=0.25$. However, the average number of iterations remains 2.6$\times$ higher than for NMS. Therefore, RHS might not provide a nominal throughput advantage on this code. However, because of the reduction in area provided by binary message passing and check node partitioning, RHS could still have an advantage in terms of throughput/area, although this would need to be confirmed at the implementation level.

\section{Conclusion}
\label{sect_conclusion}

We introduced a binary message passing decoding algorithm for LDPC codes that simplifies the wiring and the layout of fully-parallel circuit implementations, while being able to achieve an error rate that is equal to or better than the well known Sum-Product and Normalized Min-Sum algorithms.
To demonstrate the practicality of the RHS algorithm, we presented a low complexity implementation, as well as simulations results that show that the bit error rate performance remains good even at low decoding latencies.
In addition, we introduced the $\beta$-sequence method for reducing the average number of iterations with a negligible impact on implementation complexity, and we described an algorithm for resolving some decoding failures in the error floor region, named ``VN Harmonization''. This algorithm only requires additions with a pre-determined constant, operates locally in each variable node, and can be used with any BP algorithm that uses the LLR representation for variable node computations.

Our experimental results suggest that, at least for certain codes, the RHS algorithm provides a significant gain over existing algorithms in data throughput normalized to circuit area. This throughput gain can also be traded off for increased power efficiency.

\section*{Acknowledgement}
Authors wish to thank CLUMEQ and WestGrid for providing computing resources, and NSERC for funding. 

\ifCLASSOPTIONcaptionsoff
  \newpage
\fi



\bibliography{IEEEabrv,allreferences.bib}
\bibliographystyle{IEEEtran}


\end{document}